\pdfoutput=1
\documentclass[letterpaper, 10 pt, conference]{ieeeconf}  % Comment this line out if you need a4paper

\IEEEoverridecommandlockouts                              % This command is only needed if 
\usepackage{comment}
\usepackage{microtype}
\usepackage{graphicx}
\usepackage{booktabs} 
\usepackage{float}
\usepackage[utf8]{inputenc}
\usepackage{graphics,graphicx}
\usepackage{amsfonts}
\usepackage{amsmath}
\usepackage{amsthm}
\usepackage{amssymb}
\allowdisplaybreaks
\theoremstyle{definition}
\newtheorem{lemma}{\textbf{Lemma}}

\newtheorem{theorem}{\textbf{Theorem}}

\usepackage{hyperref}
\usepackage[caption = false]{subfig}
\usepackage{bbding}
\usepackage{adjustbox}
\usepackage{stix}
\usepackage{babel}
\newcommand{\HRule}{\noindent\rule{\linewidth}{0.1mm}\newline}

\renewcommand{\implies}{\;\Rightarrow\;}

\DeclareMathOperator*{\argmin}{arg\,min}

\newcommand{\sos}{\Sigma[x]}

\usepackage{color}

\usepackage{xcolor}

\usepackage[normalem]{ulem}
\usepackage[linesnumbered,ruled,vlined]{algorithm2e}
\SetKwInput{KwInput}{Input}     
\SetKwInput{KwOutput}{Output}  
\usepackage{algpseudocode}
\algnewcommand{\algorithmicand}{\textbf{ and }}
\algnewcommand{\algorithmicor}{\textbf{ or }}
\algnewcommand{\OR}{\algorithmicor}
\usepackage{stix}
\useunder{\uline}{\ul}{}

\renewcommand{\baselinestretch}{0.98}

\title{
\vspace{-2cm}
\scriptsize 2022 IEEE. Personal use of this material is permitted. Permission from IEEE must be obtained for all other uses, in any current or future media, including reprinting/republishing this material for advertising or promotional purposes, creating new collective works, for resale or redistribution to servers or lists, or reuse of any copyrighted component of this work in other works.\\ \vspace{1.5cm}
\LARGE \bf 
Koopman-based Neural Lyapunov functions for general attractors
}
\author{Shankar A. Deka, Alonso M. Valle and Claire J. Tomlin% <-this % stops a space
\thanks{
All the authors are with the Department of Electrical Engineering and Computer Sciences, University of California, Berkeley, 2594 Hearst Ave, Berkeley, CA 94720, USA. {\tt\small deka.shankar@berkeley.edu, amarco@berkeley.edu, tomlin@eecs.berkeley.edu.}
}
\thanks{This work is supported by NIFA, by the DARPA Assured Autonomy program, and by the ONR BRC program in multibody systems.}}

\begin{document}

\maketitle
\thispagestyle{empty}
\pagestyle{empty}

%%%%%%%%%%%%%%%%%%%%%%%%%%%%%%%%%%%%%%%%
\begin{abstract}
Koopman spectral theory has grown in the past decade as a powerful tool for dynamical systems analysis and control. In this paper, we show how recent data-driven techniques for estimating Koopman-Invariant subspaces with neural networks can be leveraged to extract Lyapunov certificates for the underlying system. In our work, we specifically focus on systems with a limit-cycle, beyond just an isolated equilibrium point, and use Koopman eigenfunctions to efficiently parameterize candidate Lyapunov functions to construct forward-invariant sets under some (unknown) attractor dynamics. Additionally, when the dynamics are polynomial and when neural networks are replaced by polynomials as a choice of function approximators in our approach, one can further leverage Sum-of-Squares programs and/or nonlinear programs to yield provably correct Lyapunov certificates. In such a polynomial case, our Koopman-based approach for constructing Lyapunov functions uses significantly fewer decision variables compared to directly formulating and solving a Sum-of-Squares optimization problem.
\end{abstract}

\section{Introduction}

Formal guarantees on performance and safety are important in many safety critical cyber-physical systems, like surgical systems, autonomous vehicles, or bipedal robots \cite{ames2014rapidly, amescbf}. Lyapunov-based certificates are popular, not just for studying the dynamic properties of a closed-loop system, such as convergence to a periodic orbit or existence of a unique equilibrium \cite{manchester_roa, deka2019long}, but also for the synthesis of such controllers, for instance, like in Control Lyapunov Function (CLF) \cite{ames2014rapidly} or Control Barrier Function (CBF) \cite{amescbf} based methods. % for enforcing safety.

For a general nonlinear system, converse Lyapunov theorems establish the existence of Lyapunov functions certifying various dynamical properties of the system \cite{khalil2002nonlinear,hauser_lyapunov,ahmadi2013stability}. However, these results are often not constructive in nature: they do not tell us \textit{how} to construct these Lyapunov functions. One of the computational challenges of finding a suitable candidate function lies in verifying the conditions imposed on its gradient over the entire state space or in a subset that may be of special interest. For instance, in polynomial systems, such conditions may be expressed as positive or negative semi-definiteness conditions on polynomials, the verification of which is known to be an NP-hard problem \cite{parrilo2000structured}.

A classical approach towards parameterizing these candidate functions has been the Sum-of-Squares (SOS) method \cite{parrilo2000structured}, and more recent works have focused on deploying Deep Neural Networks (DNN) to represent Lyapunov functions \cite{lyapunov_nn, formal_lyapunov,chang2019neural}. DNNs can capture intricate nonlinearities, making them powerful function approximators, but this also makes their formal verification and analysis more challenging. On the other hand, SOS-based approaches, though limited in applicability to polynomial dynamics, can provide sufficient conditions for the aforementioned semi-definiteness conditions of polynomials which are computationally tractable to verify \cite{majumdar2017funnel, singh2018robust, parrilo2000structured}.

\begin{figure}[!t]
    \centering
    \begin{picture}(100,100)
    \put(-72,0){\includegraphics[width=0.48\columnwidth]{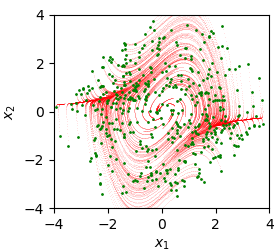}}
    \put(45,10){\includegraphics[width=0.5\columnwidth]{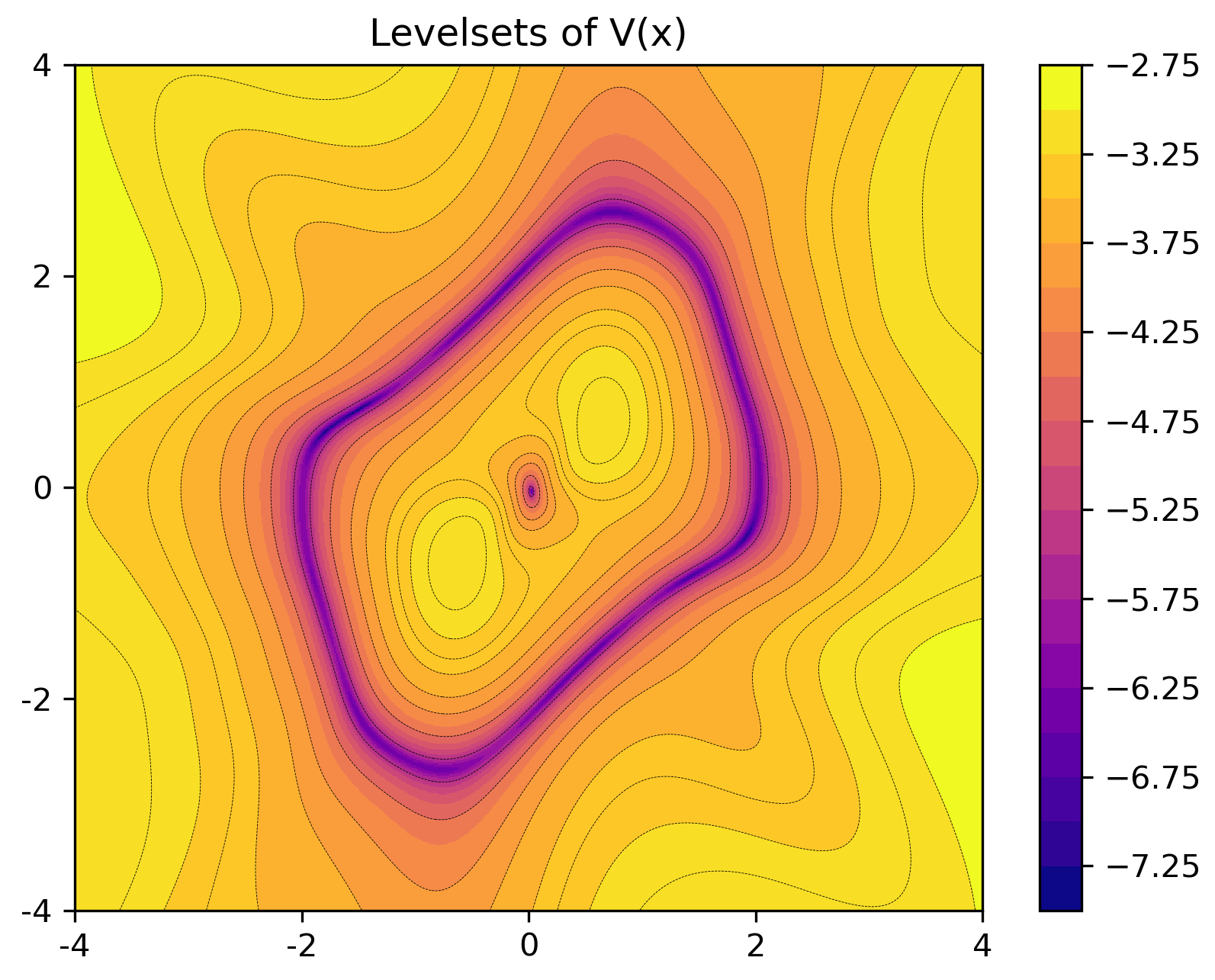}}
   \end{picture}
   \vspace{-0.3cm}
    \caption{(Left) Trajectories (in red) at various starting points (green) converge to the stable limit-cycle of the Van der Pol attractor. (Right) Contour plot of Neural Lyapunov function (in log-scale) for this system.}
    \label{fig:Lyap_initial1}
    \vspace{-0.5cm}
\end{figure}
%However, a major caveat of SOS programs is that the number of decision variables grows combinatorially with the number of state variables and the degree of polynomials. This significantly restricts their applicability because such limitations directly result in poor expressivity.

In this paper, we propose a novel methodology to utilize data-driven techniques based on Koopman operator theory \cite{brunton2021modern} to find the region of attractions (RoA) and invariant sets. Our method is applicable to systems with general attractors such as limit-cycles (Figure \ref{fig:Lyap_initial1}). %Koopman operators act linearly on the infinite dimensional space of observables.
Approximated in finite dimensions, Koopman Operators can be used to lift a nonlinear dynamical system onto a higher dimensional space where the dynamics are approximately linear. Such a ``global" linear property is shown to hold in the entire RoA \cite{williams2015data}, making it a generalized extension of the Hartman-Grobman theorem\footnote{This theorem signifies that locally around a hyperbolic equilibrium point, the flow of a dynamical system is topologically conjugate to its linearization.}
\cite{mauroy2016global}. %This property has been central to the growing popularity of Koopman-based approaches to many control system applications \cite{goswami2017global, narasingam2020data, bruder_soft_robot}. 
However, since they are obtained only in approximation using data-driven techniques \cite{williams2015data}, they cannot be directly used to construct formal certificates of the system behavior, and require further refinement. %In our work, we leverage the practicality of Koopman-based approach to construct Lyapunov certificates from trajectory data. 
In this paper, Lyapunov certificates are parameterized linearly using a finite set of learned Koopman eigenfunctions, which under certain conditions, can be verified using common optimization tools. This parameterization significantly reduces the number of decision variables.
%, as searching for the Lyapunov candidate is restricted only to the span of these Koopman eigenfunctions. 
%Moreover, approximate Lyapunov certificates constructed by our approach can in some cases, serve as a good initial guess to SOS programs that find Lyapunov functions. Since SOS program often suffers from local optima due to nonconvexity of the problem \cite{dinh_bmi}, this initial guess can significantly help the program to obtain a better solution.\\

The main contribution of our paper lies in constructing linearly parameterizable Lyapunov functions for general attractors from data, in a manner that is interpretable through the lens of Koopman Operators. Our parameterization would be ultimately beneficial if and when further optimization can be performed (like Sum-of-Squares or Nonlinear Programming in case of a known dynamics model). 
%We demonstrate our approach of combining Koopman analysis with SOS optimization in simple numerical examples. First, we show how one can obtain analytical approximations of the stable limit cycle of the Van der Pol oscillator, as well as the largest ``safe'' invariant set in the presence of obstacles. Another system we showcase is the Duffing oscillator which has a very different dynamical property. We deploy our approach to finding invariant sets that represent the local Regions of Attraction (RoAs) of the system.\\
The remainder of this paper is organized as follows. Section \ref{sec:preliminaries} provides essential background in Lyapunov based approaches for analyzing various dynamic behaviors, forward invariance in particular, and describes the preliminaries of the Koopman operator. Section \ref{sec:Neural_Lyapunov_function} contains our main approach for constructing a family of Koopman-based Lyapunov certificates from data, and an accompanying algorithm to further enforce Lyapunov constraints on the learned Lyapunov candidates. We demonstrate our approach\footnote{Code for this paper and supplementary materials can be found at: \hyperlink{https://github.com/dekovski/Koopman_Lyapunov}{https://github.com/dekovski/Koopman\_Lyapunov}} in Section \ref{sec:results} using a 2-dimensional system with a stable limit cycle, and an 11-dimensional system with multiple hyperbolic fixed points. Concluding remarks and future directions are presented in Section \ref{sec:conclusion}.

\section{Preliminaries}\label{sec:preliminaries}
Let us consider a continuous time dynamical system
\begin{eqnarray} \label{eq:dynamics}
    \frac{d}{dt}x(t) = f(x(t)),
\end{eqnarray}
where $x$ evolves in a state space $X \subset \mathbb{R}^n$. The \textit{flow map} $F^t:X\rightarrow X$ for this system is given by
\begin{equation}
\label{eq:flowmap}
 F^t(x_0) = x_0 + \int_{t_0}^{t+t_0} f(x(\tau))d\tau.
\end{equation}
A set $\mathcal{W}\subseteq X$ is \textit{forward-invariant} if for every $t>0$ and $x_0 \in \mathcal{W}$, we have $F^t(x_0)\in\mathcal{W}$. We can now present some essential Koopman Operator preliminaries.

Given the \textit{space of observables} $\mathcal{F}$, defined as the set of all observable functions mapping $X\rightarrow \mathbb{C}$, the Koopman operator $\mathcal{K}^t: \mathcal{F} \rightarrow \mathcal{F}$ is then defined as an operator acting on an observable $g:X \rightarrow \mathbb{C}$ in $\mathcal{F}$ such that
\begin{eqnarray*}
    \mathcal{K}^t g = g\circ F^t.
\end{eqnarray*}
An eigenfunction $\psi \in \mathcal{F}$ of the Koopman operator $\mathcal{K}^t$ satisfies
\begin{eqnarray*} 
    \mathcal{K}^t \psi &=& e^{\lambda t} \psi \\ \nonumber
    \frac{d}{dt} \psi(x(t)) &=& \lambda \psi(x(t)),\nonumber
\end{eqnarray*}
for some $\lambda \in \mathbb{C}$. Although Koopman operators are infinite dimensional linear operators, one may obtain finite dimensional approximations through data-driven approaches like the Extended Dynamic Mode Decomposition (EDMD) \cite{williams2015data}. Let $\mathcal{F}_N \subset \mathcal{F}$ be a $N$-dimensional \textit{Koopman Invariant Subspace}, that is, for any $g \in \mathcal{F}_N$, we have $\mathcal{K}^tg \in \mathcal{F}_N$. Next, we consider function $\Phi(x)=\left[\phi_1(x),\phi_2(x),\ldots,\phi_N(x)\right]^\top$ comprised of basis functions $\phi_i(x) \in \mathcal{F}_N, \; i=1,2,..,N$.\\
Given that we have trajectory snapshots in form of $M$ pairs $(x_i,y_i)$ where $y_i = F^t(x_i)$ for $i=1,2,...,T$, the EDMD procedure is used to estimate the Koopman matrix $K$ by solving the following least-squares problem:
\begin{equation}\label{eq:EDMD}
K = \underset{A \in \mathbb{C}^{N \times N}}{\argmin} \|\Phi(Y) - A\Phi(X)\|^2_F %= \underset{A \in \mathbb{C}^{N \times N}}{\argmin} \sum_{i=1}^{M} \|\Phi(y_i) - A\Phi(x_i)\|^2_2,
\end{equation}
where $\|\cdot\|_F$ denotes the Frobenius norm, and matrices $\Phi(X) = [\Phi(x_1), \Phi(x_2), ..., \Phi(x_T)]$ and $\Phi(Y) = [\Phi(y_1),\Phi(y_2),...,\Phi(y_T)]$. The $K$ that minimizes \eqref{eq:EDMD} is obtained in closed-form as
\begin{eqnarray*}
    K = \Phi_{XY}\Phi_{XX}^{\dagger},
\end{eqnarray*}
where $\dagger$ denotes the psuedo inverse, $\Phi_{XY} = \Phi(Y)\Phi(X)^T$ and $\Phi_{XX} = \Phi(X)\Phi(X)^T$. A popular choice of the basis (or dictionary) functions $\Phi$ are monomials. Other choices may include radial basis functions \cite{williams2015data}, deep neural network-based functions \cite{lusch2018deep}, and Taylor or Bernstein polynomials \cite{mauroy2016global}.\\
 Suppose that $(v_i,\mu_i)$ are the eigenvectors and eigenvalues of $K^T$ for $i=1,2,..,M$, then it is straightforward to show:
\begin{eqnarray}\label{eq:eigenfunction}
 \psi_i(x) :=  v_i^T\Phi(x)
\end{eqnarray}
is a Koopman eigenfunction, with eigenvalue $\lambda_i = \frac{1}{t}\log(\mu_i)$. 
Such a spectral decomposition of the Koopman operator yields eigenfunctions, which, along with their eigenvalues, contain rich information about the underlying system dynamics \cite{mauroy2016global}. These serve as a basis building block for our Koopman-Lyapunov functions. Although EDMD is a practical technique for estimating these eigenfunctions, it solves the minimization \eqref{eq:EDMD} over just one timestep, which can often lead to approximation errors accumulating over larger time horizons. One contribution of our paper is to utilize deep learning to perform a multi-timestep optimization for more accurate estimation of Koopman eigenfunctions, as presented in Subsection \ref{subsec:Learning Koopman}
\section{Neural-Lyapunov function}\label{sec:Neural_Lyapunov_function}

This is the main section of our paper, where we build Koopman-Lyapunov functions and construct a family of candidates that under perfect learning, meaning no approximation errors, characterize invariant sets and RoA for stable limit sets. We then present a sampling based algorithm that utilizes convex polytopes to further eliminate infeasible Lyapunov candidates arising due to approximation errors. 
\subsection{Learning Koopman Operator and Eigenfunctions}\label{subsec:Learning Koopman}
\begin{figure}[!htp]
    \centering
    \includegraphics[width=\columnwidth]{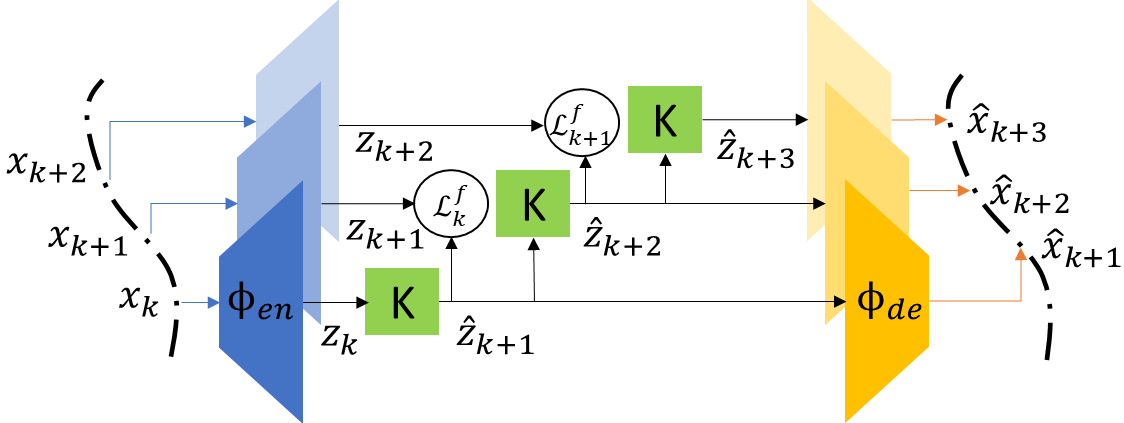}
    \caption{Neural network architecture for multi-timestep optimization.}
    \label{fig:Network}
\end{figure}

In this section, we present a multi-timestep optimization to estimate the Koopman operator, inspired by \cite{azencot2020forecasting}. Towards that end, we use a neural network to learn the basis functions alongside the Koopman matrix, as depicted in Figure \ref{fig:Network}. Once we lift our dynamics to this Koopman space, we extract our neural-Lyapunov functions, which we shall discuss in the next subsection. As opposed to other works that directly parameterize the Lyapunov function as a neural network, our approach \textit{implicitly} learns it, as a by-product of learning the Koopman lift and the stable eigenfunctions. This is done primarily to obtain a linearly parameterized family of possible Lyapunov candidates, which is very beneficial when fine-tuning these functions further. This also adds to the interpretability of our neural net model as opposed to training a network to directly learn a Lyapunov function. %\textcolor{red}{Moreover, while a Lyapunov function may or may not exist for an underlying dynamics (For example, when the limit-cycle or fixed-point is unstable), eigenfunctions can always be defined (atleast locally around the limit-cycle or fixed-point). Therefore, learning the Koopman lifts is a more well-defined problem than directly attempting to learn a Lyapunov function.}

We adopt an encoder-decoder architecture to perform the Koopman lift $x \mapsto \Phi_{en}(x) $ and then transform the lifted space back to the original state-space, $z \mapsto \Phi_{de}(z)$. The associated loss function for this autoencoder part of the network is given by
\[
\mathcal{L}^{ae}(x) = \|x - \Phi_{de}\circ\Phi_{en}(x)\|^2_2.
\] 
Given trajectory snapshots $x_{k:k+T} = \left[ x_k,x_{k+1},\ldots,x_{k+T}\right]$at uniform time intervals, the Koopman matrix $K$ is obtained by minimizing the $T$-timestep forward prediction loss function:
\[
\mathcal{L}^f(x_{k:k+T}) = \sum_{i=0}^{T-1} \mathcal{L}^f_{k+i} = \sum_{i=0}^{T-1} \|\Phi_{en}(x_{k+i+1}) - K^{i+1}\Phi_{en}(x_{k})\|^2_2.
\] 
Note that in contrast to the EDMD optimization problem described by equation \eqref{eq:EDMD}, the minimizer of $\mathcal{L}^f$ cannot be obtained in closed-form (except when $T=1$). This added cost of optimization is well justified by a more precise estimation of the Koopman operator than EDMD \cite{azencot2020forecasting}. Put together, we solve 
\begin{eqnarray}
    \min_{\theta,K} \; \underset{x_{0:T} \sim X_{data}}{E} \left[ \frac{1}{T}\sum_{i=0}^{T-1}\left(p_1\cdot\mathcal{L}^{ae}(x_{i}) + p_2\cdot\mathcal{L}^f_{i}(x_{0:T})\right)\right],
\end{eqnarray}
where $\theta$ denotes the parameters of the encoder-decoder network, and $p_1,p_2$ are positive hyperparameters.

Although Koopman representations are typically used for forecasting trajectories in either the lifted Koopman space or the original state-space (shown in Figure \ref{fig:Network} respectively as $\hat{z}$ and $\hat{x}$), we are primarily interested in the Koopman eigenfunctions and their utility in constructing Lyapunov functions, which we consider next.
\subsection{Linear space of Lyapunov candidates}
Let us consider a set of eigenfunctions 
\begin{eqnarray*}
\Psi = \big\{\psi_i \in \mathcal{F} \; \rvert \; \Re(\lambda_i) < 0, \, i=1,2,...,M \big\},
\end{eqnarray*} with eigenvalues $\lambda_i \in \mathbb{C}$. If we define $V_i \doteq \frac{1}{2}\|\psi_i\|^2$ for each $i=1,2,...,M$, then $V_i$ is a Lyapunov function, satisfying $V_i(x) \ge 0, \dot{V}_i(x) \le 0$ for all $x \in X$. Thus, any sub-level set 
\begin{eqnarray*}
\mathcal{M}_i^c \doteq \left\{ x \; \vert \; V_i(x)\le c  \right\}
\end{eqnarray*}
where $c$ is a non-negative constant, will be forward invariant. Additionally, when $\lambda_i < 0$, the zero-level set $\mathcal{M}_i^0$ is globally asymptotically stable \cite{mauroy2016global}. As a corollary, the following set is forward invariant and globally asymptotically stable: 
\begin{eqnarray*}
    \mathcal{M} = \bigcap \limits_{i=1}^M \mathcal{M}_i^0.
\end{eqnarray*} 
We then construct Lyapunov functions from these $V_i$'s by simply taking their weighted sum:
\begin{eqnarray}\label{eq:Lyapunov}
 \mathcal{V}_{Lyap} = \left\{ \sum_{i=1}^{M} a_iV_i(x) \; \vert \;  a_1, a_2, \ldots , a_M > 0.\right\}
\end{eqnarray}
More generally, since the any finite product of Koopman eigenfunctions is also an eigenfunction, it is easy to see that the following function can also serve as a Lyapunov candidate:
\begin{eqnarray}
\tilde{V}(x) = \sum_{i=1}^{M} a_iV_i + \sum_{i,j=1}^{M} a_{ij}V_iV_j + \sum_{i,j,k=1}^{M} a_{ijk}V_iV_jV_k + ...
\end{eqnarray}
It follows from equation \eqref{eq:Lyapunov} that any function $V(x)\in \mathcal{V}_{Lyap}$ constructed using stable polynomial eigenfunctions belongs to the set of sum-of-squares (SOS) polynomial, which we denote by $\sos$. \textit{Ideally}, this $V(x)$ would be a Lyapunov function with $\dot{V}(x) \le 0$ along the trajectories of the system \eqref{eq:dynamics}. Unfortunately, the negative semi-definiteness condition may not actually hold, since EDMD yields only an approximation of the Koopman operator and its eigenfunctions. However, if the dynamics \eqref{eq:dynamics} is polynomial, then one may setup a SOS optimization problem to find a correct Lyapunov function through EDMD, as we show later in Section \ref{sec:results}.

Let us again consider the space of candidate functions \eqref{eq:Lyapunov} described by the finite dimensional linear space spanned by $V_i$ for $i=1,2,...,M$ obtained from the learned Koopman eigenfunctions. Then, for each $i$, we have
\begin{eqnarray}\label{eq:error}
    \dot{V}_i = \lambda_iV_i(x) + \varepsilon_i(x),
\end{eqnarray}
where $\varepsilon_i(x)$ accounts for the approximation error in our learning process. Under assumptions on the boundedness of these approximation errors, we can still provide Lyapunov-based guarantees on the system using elements of $\mathcal{V}_{Lyap}$, as stated in the following. 
\begin{theorem}\label{th:main}
Let us assume that the approximation errors is bounded for each $i$ as $\|\varepsilon_i(x)\| \le \kappa_i\|V_i(x)\|^2 + \omega_i$ for some positive constants $\kappa_i$ and $\omega_i$. Then, there exists a function $V\in \mathcal{V}_{Lyap}$ described by equation \eqref{eq:Lyapunov} and a scalar $\gamma>0$ such that the $\gamma$-sublevel of $V(x)$ is forward invariant, for sufficiently negative eigenvalues $\lambda_i, \; i=1,2,\ldots,M$.
\end{theorem}
\noindent\textit{Proof.} Please see Appendix \ref{sec:AppendixA}.\\

The boundedness assumption for $\varepsilon_i(x)$ in Theorem \ref{th:main} is a mild one, and can be shown to hold if $(\nabla^\top \psi_i) f$ is bounded. We present this as the following lemma.

\begin{lemma}\label{lemma:bounded}
Given the approximation error $\varepsilon_i$ in the Lyapunov basis $V_i(x)$ described by equation \eqref{eq:error}, if $\omega_i$ and $\kappa_i$ satisfy
\[
\omega_i > \frac{(p+\lambda_i)^2}{4\kappa_i} - \frac{q^2}{4p^2}
\]
for some constants $q>\|(\nabla^\top\psi_i)f\|$ and $p>0$, then $\|\varepsilon_i\| \le \kappa_i\|V_i(x)\|^2 + \omega_i$.
\end{lemma}
\noindent\textit{Proof}. Please see Appendix \ref{sec:AppendixB}.

\subsection{Set of feasible candidate Lyapunov functions as polytopes}

In this subsection, we present a sampling-based algorithm to find a set of \textit{feasible} Lyapunov candidates from the finite dimensional, linear space of functions $\mathcal{V}_{Lyap}$ given by equation \eqref{eq:Lyapunov}. If the Koopman eigenfunctions $\phi_i$'s are known exactly, every element in the set $\mathcal{V}_{Lyap}$ is a Lyapunov function by construction. However, in practice, these $\phi_i$'s are learned from data, as described in the previous sections. Inevitably, the approximation errors in $\phi_i$'s and consequently $V_i$'s, may lead to violation of the negative definiteness condition, $\dot V <0$, for some $V(x) \in \mathcal{V}_{Lyap}$. Fortunately, our linear parameterization allows us to efficiently eliminate infeasible Lyapunov candidates in the set $\mathcal{V}_{Lyap}$ by enforcing negative semi-definiteness Lyapunov conditions on the elements $V\in\mathcal{V}_{Lyap}$ over sampled data-points. Such a sampling-based approach is particularly useful for cases when the intricate nonlinearities in $V(x)$ prevent analytical or optimization-based verification of Lyapunov conditions - a problem commonly faced when employing deep neural networks as function approximators.

\begin{figure}[!t]
    \centering
    \vspace{0.3cm}
    \includegraphics[width=\columnwidth]{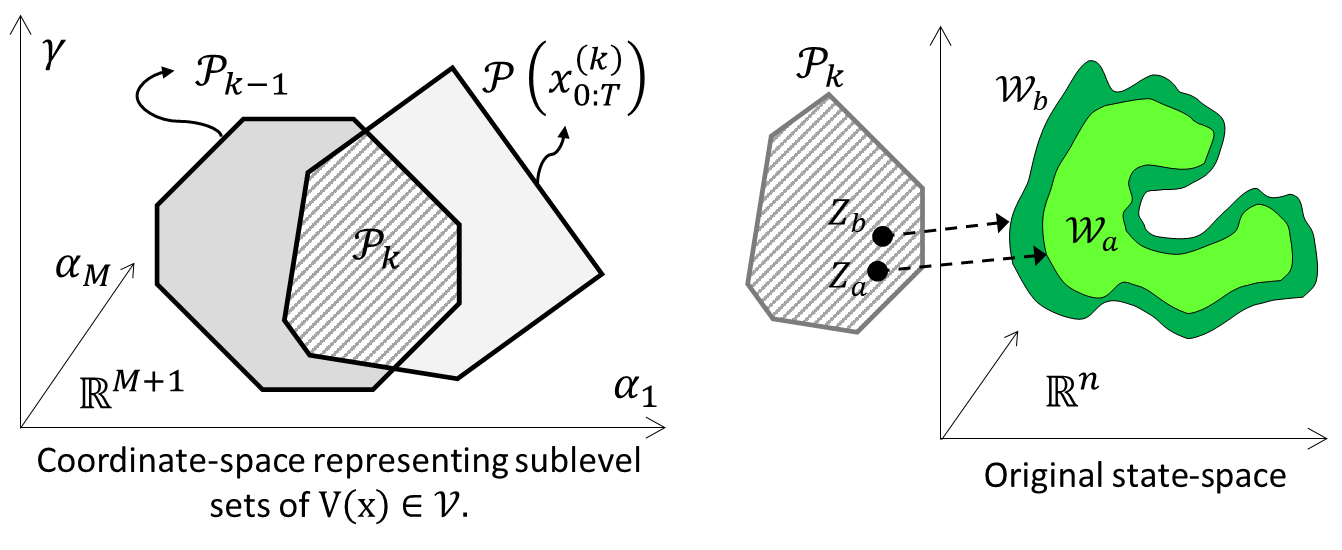}
    %\vspace{-0.3cm}
    \caption{Feasible Lyapunov candidates from the set $\mathcal{V}_{Lyap}$ are described as convex polytopes that are refined iteratively. Each point in the polytope $Z$ represents a region $\mathcal{W}$ in the state-space that encloses the set $\mathcal{U}$ and is a candidate invariant set.}
    \label{fig:Polytope}
\vspace{-0.3cm}
\end{figure}

In order to find a Lyapunov function with some $\gamma$-sublevel set that is forward invariant and contains some set $\mathcal{U}$, we would like the following sufficient condition to hold:
\begin{eqnarray}\label{eq:Lyapunov_condition}
     x \in \mathcal{U} \subseteq \left\{ x \vert V(x) \le \gamma \right\} \implies \dot{V} \le \beta\left(\gamma - V(x)\right)
\end{eqnarray}
Let us say we sample trajectory $x_{0:T} = \left[x_0,x_1,\ldots,x_T\right]$ from $\mathcal{U}$. %, sampled at sufficiently small timestep $dt$, we approximate the term $\dot{V}$ at some point $x_k$ as the discrete difference $\frac{1}{dt}\left(V(x_{k+1}) - V(x_k)\right)$. 
Then, for $V$ in our linear space of Lyapunov candidates $\mathcal{V}_{Lyap}$, we must satisfy
\begin{align}
\begin{split}
    A_1Z &\le 0, \\
    \left(A_2 + \beta A_1\right)Z &\le 0,\\
    Z &\ge 0,
\end{split}
\label{eq:polytope}
\end{align}
\begin{gather*}
    \text{where } A_{1} \doteq
    \left[\begin{matrix}
        V_1(x_0) & V_2(x_0) & \cdots & V_M(x_0) & -1\\
        V_1(x_1) & V_2(x_1) & \cdots & V_M(x_1) & -1\\
        \vdots & \vdots & \ddots & \vdots  & \vdots\\
        V_1(x_T) & V_2(x_T) & \cdots & V_M(x_T) & -1\\
    \end{matrix}\right],\\
    \\
    A_{2} \doteq
    \left[\begin{matrix}
        \dot{V}_1(x_0) & \dot{V}_2(x_0) & \cdots & \dot{V}_M(x_0) & 0\\
        \dot{V}_1(x_1) & \dot{V}_2(x_1) & \cdots & \dot{V}_M(x_1) & 0\\
        \vdots & \vdots & \ddots & \vdots  & \vdots\\
        \dot{V}_1(x_T) & \dot{V}_2(x_T) & \cdots & \dot{V}_M(x_T) & 0\\
    \end{matrix}\right],\; Z \doteq
    \left[\begin{matrix}
        \alpha_1 \\
        \alpha_2 \\
        \vdots \\
        \alpha_M \\
        \gamma \\
    \end{matrix}\right].
\end{gather*}
Equation \eqref{eq:polytope} represents a convex polytope $\mathcal{P}(x_{0:T})$ containing a set of positive coefficients $\alpha_1,\alpha_2,\ldots,\alpha_M$ and a corresponding positive scalar $\gamma$ that together describe a candidate invariant set of the form $\big\{x \; \vert \; \sum_{i=1}^M \alpha_i V_i(x) \le \gamma \big\}.$ If we sample multiple trajectories, the polytope $\mathcal{P}$ can be refined iteratively, by taking intersection $\mathcal{P}_k = \bigcap_{i=0}^k \mathcal{P}(x_{0:T}^{(k)})$, where $x_{0:T}^{(k)}$ denotes the trajectory data sampled at the $k^{th}$ iteration. This idea is illustrated by Figure \ref{fig:Polytope}, and presented concisely in Algorithm \ref{algo:Polytope}.
%\vspace{0.3cm}

By construction, $P_k$ converges to some convex polytope $P_\infty$ as $k\rightarrow \infty$, since its volume decreases monotonically with the number of iterations. If the set $\mathcal{U}$ lies inside the $\gamma$-sublevel set of the function $V(x)=\sum_{i=1}^M \alpha_iV_i(x)$ described in Theorem \ref{th:main}, then the $\left[\alpha_1,\ldots,\alpha_M,\gamma\right]\in P_\infty$.

\begin{comment}
As $T\rightarrow \infty$, if this limit $\mathcal{P}_\infty \neq \emptyset$ one can show that the set
\begin{eqnarray*}
 \bigcap_{\alpha,\gamma \in \mathcal{P}_\infty} \big\{
x \big\vert \sum_i \alpha_iV_i(x) < \gamma
\big\}
\end{eqnarray*}
bounds the smallest invariant set containing $\mathcal{U}$.
\end{comment}
\vspace{0.1cm}
\SetNlSty{text}{}{:}
\SetKw{KwBy}{by}
\begin{algorithm}[!htp]
\SetAlgoLined
\vspace{0.1cm}
 \KwInput{Koopman-Lyapunov basis functions $V_1,V_2,\ldots,V_M$, system dynamics, set $\mathcal{U}$, positive scalar $\beta$, maximum iterations \textsc{MAXITER}, number of trajectory samples \textsc{N} and trajectory length \textsc{T}.}
 \KwOutput{Coefficients \textbf{$\alpha_1$},\textbf{$\alpha_2$},$\dots$,\textbf{$\alpha_M$} corresponding to Koopman-Lyapunov basis functions, and levelset value $\gamma$ describing invariant set.}
 Initialize: $ \mathcal{P} \gets \big\{Z \; \vert \;  Z \in [0,1]^{M+1}, \sum_{i=1}^M Z_i \ge 1 \big\}$,  $k \gets 0$. \\
 \While{{k<MAXITER}}{
 $X,Y \gets \text{SampleData}(\mathcal{U});$\\
 $V_X \gets \big[V_{basis}(X)\; , \;-\mathbf{1}\big], \;V_Y \gets \big[V_{basis}(Y) \; , \;-\mathbf{1}\big]$;\\
 $V_{dot} \gets  \frac{1}{dt}\left(V_{basis}(Y) - V_{basis}(X)\right)$;\\ 
 $\mathcal{P}'\gets \big\{Z \; \vert \; V_XZ \le 0 \; , \; (V_{dot} + \beta V_X)Z \le \mathbf{0} \big\} \bigcap \mathcal{P}; $\\
 \If{$\mathcal{P}' = \emptyset$}{
    break
 }
    $\mathcal{P} \gets \mathcal{P}';$\\
    $k \gets k+1$
 }
\nl $x \sim \mathcal{U}^c;$ \tcp{Sample point outside set U.}
\nl $\mathcal{P} \gets  \big\{Z \; \vert \; V_{basis}(x)Z > 0\big\} \bigcap \mathcal{P}$ \\
\nl\KwRet{\text{Polytope} $\mathcal{P}$}
\caption{Invariant set computation using Koopman-Lyapunov functions}
\label{algo:Polytope}
\end{algorithm}
\vspace{-0.2cm}
\section{Numerical Results}\label{sec:results}
We first consider the two-dimensional Van der Pol oscillator system with a stable limit cycle, which has a polynomial right-hand side, given by
\begin{eqnarray*}
    \dot{x}_1 &=& x_2\\
    \dot{x}_2 &=& -x_1+x_2(1-x_1^2).
\end{eqnarray*}
The Koopman lift is obtained using the neural network architecture described in Section \ref{sec:Neural_Lyapunov_function}, with $\Phi_{en}$ and $\Phi_{de}$ both taken as a $4$-layer feedforward network with $\tanh$ activation. The output layer of $\Phi_{en}$ is of size $20$, which is the chosen dimension of the Koopman space. Of these 20 Koopman basis functions, $M=18$ stable eigenfunctions were then constructed using equation \eqref{eq:eigenfunction}.

\begin{figure}[!t]
\subfloat[]{
\begin{minipage}{0.49\columnwidth}
\centering
    \includegraphics[width=\columnwidth]{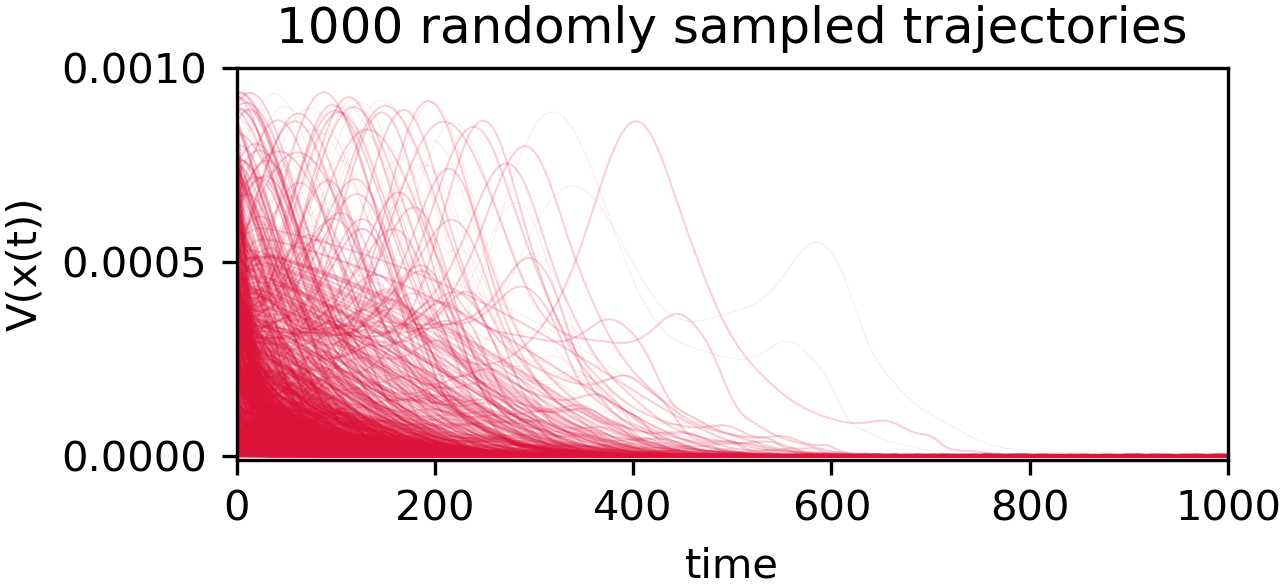}
    %\vspace{-0.1cm}
    \includegraphics[width=\columnwidth]{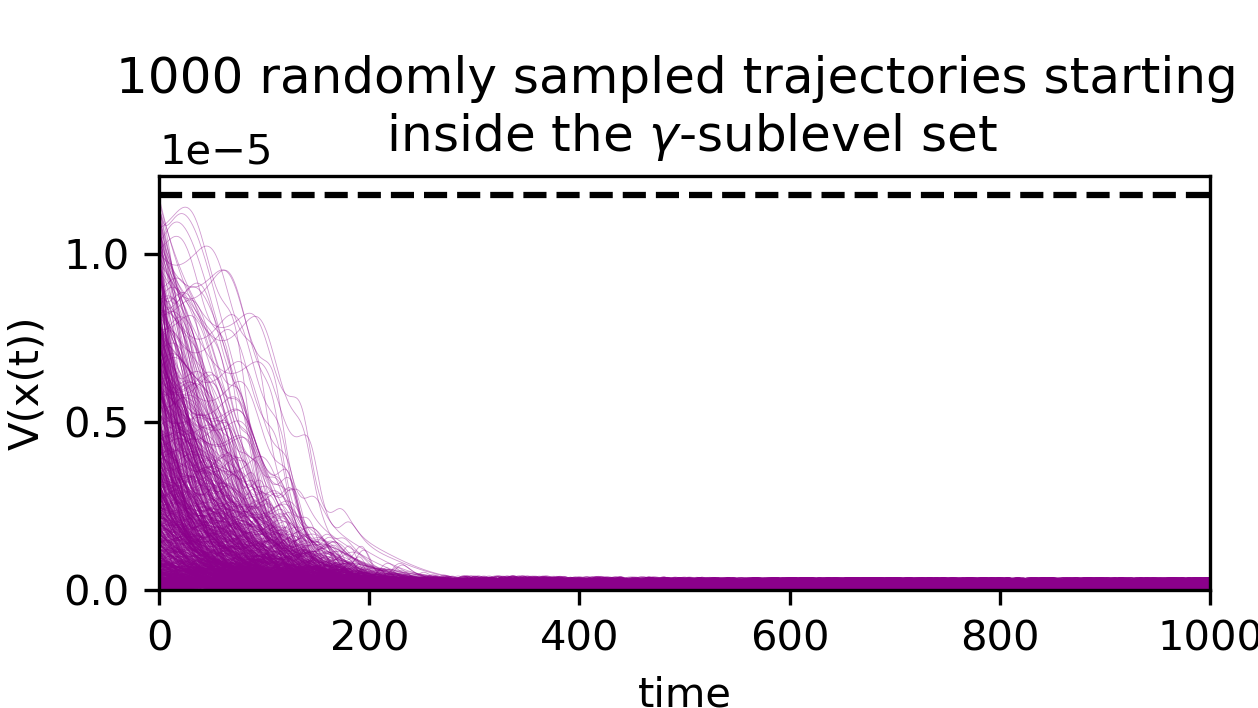}
\end{minipage}}
\subfloat[]{
\begin{minipage}{0.49\columnwidth}
    \centering
    \vspace{0.2cm}\includegraphics[width=\columnwidth]{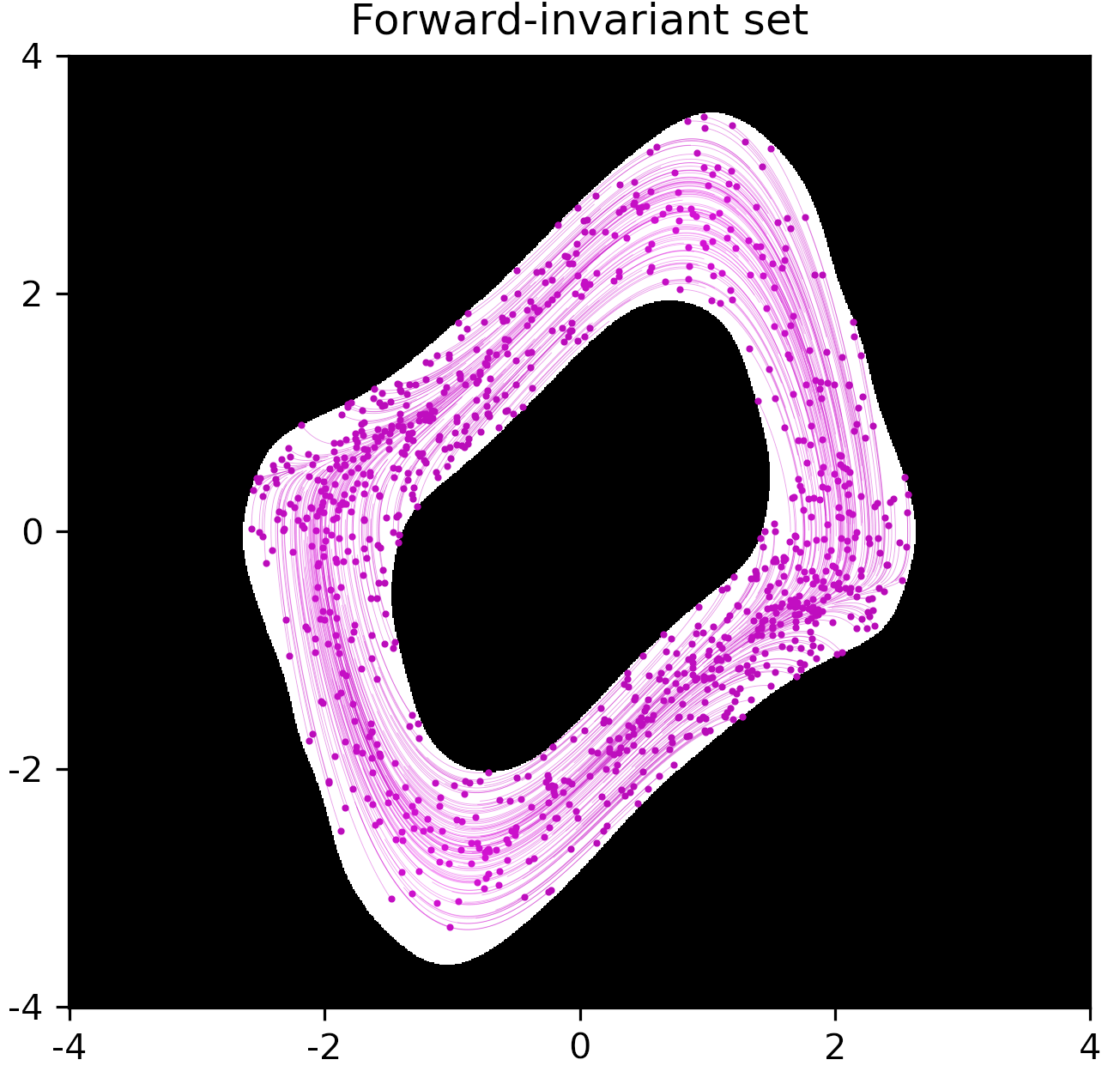}
    %\vspace{-0.1mm}
\end{minipage}}
    %\caption{$\gamma = 1.174 \times 10^{-5}, \; \alpha = [0.7431 , 0.00287, 0.001  , 0.001  , 0.001  , 0.001  ,$ $ 0.00169, 
    %   0.     , 0.001  , 0.     , 0.00026, 0.     , 0.00002, 0.     ,
   %   0.     , 0.00491, 0.001  , 0.24943]$}
  \caption{(a) Candidate functions in set $\mathcal{V}_{Lyap}$ may violate Lyapunov conditions (red). Algorithm \ref{algo:Polytope} helps refine this set $\mathcal{V}_{Lyap}$ to obtain a smaller set of feasible Lyapunov candidates for establishing forward-Invariance (bottom). (b) White region shows the $\gamma$-sublevel set ( $\gamma=1.174$) of a function $V(x)$ found after applying Algorithm \ref{algo:Polytope}, wherein trajectories starting inside remain inside (purple).}
  \label{fig:Lyap_invariant}
\vspace{-0.3cm}
\end{figure}

Figure \ref{fig:Lyap_initial1} on page 1 visualizes a Lyapunov candidate function from the set $\mathcal{V}_{Lyap}$, with $V_i$'s constructed using the estimated stable Koopman eigenfunctions. The weights $\alpha_i$'s are set as $\alpha_i = \text{softmax}(-\hat{\epsilon}_i) = \exp({-\hat{\epsilon}_i})/\sum_i \exp({-\hat{\epsilon}_i})$ based on the empirically obtained bounds $\|\epsilon_i\| \le \hat{\epsilon}_i$. Such a choice of weights ensures that the Lyapunov basis functions that are estimated more precisely, dominate $V(x) = \sum_i \alpha_iV_i(x)$. However, such a heuristic choice of $\alpha_i$'s is still not enough. When this $V(x)$ is evaluated along randomly sampled trajectories (shown in Figure \ref{fig:Lyap_invariant}(a) top panel), they do not decrease monotonically with time, due to the inevitable approximation errors in constructing $V(x)$ from data. After we apply our Algorithm \ref{algo:Polytope}, the resulting Lyapunov function accurately characterizes a forward invariant set, as shown by the white annular region in Figure \ref{fig:Lyap_invariant}. As shown in Figure \ref{fig:Lyap_invariant} (a) and (b), the trajectories in purple randomly sampled within this set stay inside this set.

\subsection{Formal verification of polynomial Lyapunov functions}
Next, we demonstrate how our data-driven, Koopman-based construction of Lyapunov functions can be verified to be correct via numerical optimization techniques, when 1) the dynamics of the underlying system is known (and polynomial), and 2) monomial basis are used in the construction of the lifted Koopman-space (meaning that the Koopman eigenfunctions are polynomials).

We first consider SOS programming, wherein the negative-definiteness condition of equation \eqref{eq:Lyapunov_condition} is established by finding a feasible solution to the following problem. Note that such a SOS feasibility is sufficient but not necessary for equation \eqref{eq:Lyapunov_condition} to hold. We use the Matlab optimization toolbox, Yalmip \cite{Lofberg2004}, with Mosek solver \cite{aps2019mosek} to solve this program.

%\filbreak
\begin{center}
\HRule
I. Sum-of-Square programming (SOS)\\
\vspace{-0.2cm}
\HRule
\vspace{-0.5cm}
\end{center}
\begin{align*} \label{eq:SOS_gamma}
\begin{split}
 \textbf{Find: }& s(x) \in \Sigma[x]\\
 \textbf{Such that: }& -(L_fV(x)-\beta(\gamma - V(x)))\\ 
 &- s(x)(\gamma - V(x)) \in \Sigma[x]
\end{split}
\end{align*}
\begin{center}
\HRule
\end{center}
Another alternative for verifying the Lyapunov certificates obtained via our data-driven approach, is to directly search for a point within the domain of interest (defined by the $\gamma$-sublevel set of $V(x)$), that violates the Lyapunov condition. Clearly, if the following maximum value is negative, then our Lyapunov certificate is verified (we use SciPy.optimize package to solve this \cite{2020SciPy-NMeth}).
\begin{center}
%\line(1,0){\columnwidth}\\
\HRule
II. Nonlinear programming (NLP)
\vspace{-0.2cm}\\
\HRule
\end{center}
\vspace{-0.5cm}
\begin{align*}
\begin{split}
 \textbf{Maximize: }& L_fV(x)-\beta(\gamma - V(x))\\
 \textbf{Such that: }& V(x) \le \gamma
\end{split}
\end{align*}
\begin{center}
\HRule
\end{center}

\begin{figure}[!t]
\vspace{0.4cm}
\begin{minipage}{0.45\columnwidth}
\includegraphics[width=\columnwidth]{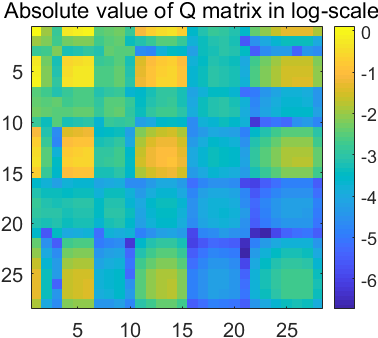}
\end{minipage}\;
\begin{minipage}{0.5\columnwidth}
\centering
$V(x) = m(x)^TQm(x), \; Q\succ 0,$\\
$\gamma = 0.0672, \beta = 1$.\\
\vspace{0.3cm}
\resizebox{\columnwidth}{!}{%
\begin{tabular}{|c|c|c|}
\hline
Verification: & SOS & NLP \\ \hline
Successful:      & yes & yes \\ \hline
\end{tabular}%
}
\end{minipage}
\caption{Polynomial Lyapunov function after constructing the linear space $\mathcal{V}_{Lyap}$ using Koopman eigenfunctions followed by Algorithm \ref{algo:Polytope}. For more, details please refer to the paper code and data. The invariant set obtained is comparable to Figure \ref{fig:Lyap_invariant}(b) and hence is not shown here again.}
\label{fig:Polynomial_lyap}
\vspace{-0.3cm}
\end{figure}

We choose monomials of maximum degree $6$ in states $x_1$ and $x_2$, leading to a total of $28$ monomial basis terms $m(x) = \left[1,x_1,x_2,x_1^2,x_1x_2,\ldots,x_1x_2^5,x_2^6\right]$. After the EDMD process, we pick $4$ stable eigenfunctions $\Psi_i(x)$ and apply our procedure to find a Lyapunov function $V(x)$ and its corresponding invariant sublevel set, shown in Figure \ref{fig:Polynomial_lyap}.

\subsection{Higher-dimensional systems}
\begin{figure}[!htp]
\centering
\vspace{0.2cm}
\hspace{0.8cm}\includegraphics[width=0.8\columnwidth]{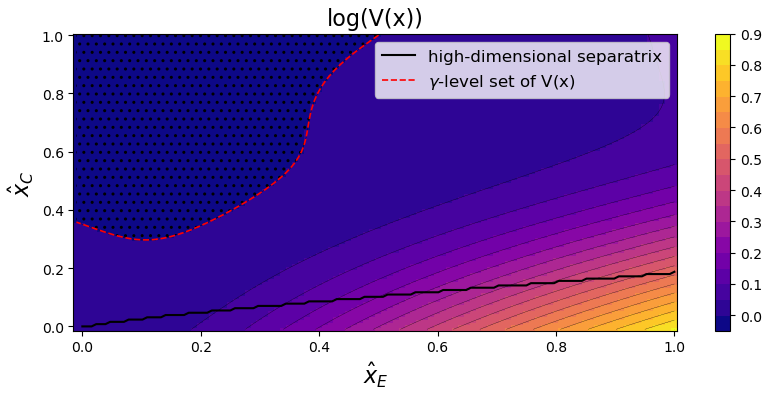}\\
\begin{picture}(100,100)
\put(-40,0){\includegraphics[width=0.8\columnwidth]{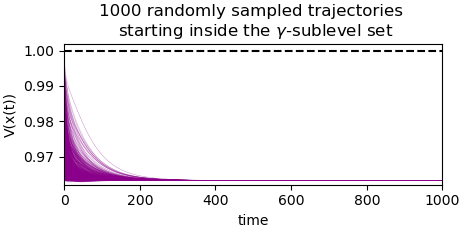}}
\begin{comment}
\put(20,60){$V(x) = m(x)^TQm(x), \; Q \succ 0,$}\\
\put(40,50){$\gamma = 0.9999, \beta = 10.$}\\
\put(30,32){
\centering
\resizebox{0.45\columnwidth}{!}{%
\begin{tabular}{|c|c|c|}
\hline
Verification: & SOS & NLP \\ \hline
Successful: & $-$ & yes \\ \hline
\end{tabular}}
}
\end{comment}
\end{picture}
\begin{gather*}
	V(x) = m(x)^TQm(x), \; Q \succ 0,\\
	\gamma = 0.9999, \beta = 10.
\end{gather*}
\centering
\resizebox{0.5\columnwidth}{!}{%
\begin{tabular}{|c|c|c|}
\hline
Verification: & SOS & NLP \\ \hline
Successful: & $-$ & yes \\ \hline
\end{tabular}}\\

\caption{(Top) Invariant set for equilibrium point `C' shown by the dotted region obtained by our approach. Lyapunov function $V(x)$ in 11-state variables is visualized in a scaled, two-dimensional plane, where points $(1,0), (0,1)$ and $(0,0)$ respectively correspond to equilibrium points `C' (healthy microbial state), `E' (antibiotics-depleted state), and the origin of the original system in $\mathbb{R}^{11}$ (See \cite{jones2019steady} for details on this 2-d projection technique). The high-dimensional separatrix (solid black) obtained by numerical simulation \cite{jones2019steady} separates the trajectories starting on the 2-d plane moving towards either equilibrium `C' or `E'. (Middle) Random trajectories uniformly sampled inside the $\gamma$-levelset are shown to stay within this levelset. (Bottom) Lyapunov function is a SOS polynomial by construction, with $m(x)$ depicting monomials of maximum degree 3 in 11 variables. Verification via SOS was found intractable.}
\label{fig:Trajs_gLV}
\end{figure}

A major concern that arises while constructing any kind of certificate for describing the dynamic behavior of a system (such as Lyapunov, Barrier, or Reachability-based) is extension and applicability to high dimensional systems. In this subsection, we consider an 11-dimensional generalized Lotka-Volterra (gLV) model used to study microbial interactions in mice-gut microbiome \cite{jones2019steady}. The gLV equations modeling $N$ interacting species within an ecological system is given by:
\begin{gather*}
    \frac{d}{dt} x_i(t) = x_i(t)\left(\rho_i + \sum_{j=1}^{N} K_{ij}x_j(t)\right), \text{for }i=1,2,\ldots,N.
\end{gather*}
The specific gLV parameter values ($\rho_i,K_{ij}$) for mice-gut microbiome were obtained experimentally and can be found in \cite{stein2013ecological}. We pick a stable equilibrium point, which is labelled as `C' in \cite{jones2019steady} and corresponds to ``healthy'' microbial population. We construct an invariant set containing this equilibrium point `C'. We use monomials again, with maximum degree $3$ in $11$ state variables to obtain $364$ Koopman eigenfunctions from data, of which we choose top $M=200$ stable eigenfunctions to construct space $\mathcal{V}_{Lyap}$. Subsequently, we use Algorithm \ref{algo:Polytope} to obtain suitable elements within $\mathcal{V}_{Lyap}$ that satisfy Lyapunov condition \eqref{eq:Lyapunov_condition} at data points sampled uniformly randomly from an $\epsilon-$ball centered at equilibrium `C' (with $\epsilon$=5). This Lyapunov function and its corresponding invariant $\gamma$-levelset is visualized in Figure \ref{fig:Trajs_gLV}.

\begin{figure}[!htp]
    \centering
    \hspace{0.5cm}\includegraphics[width=0.8\columnwidth]{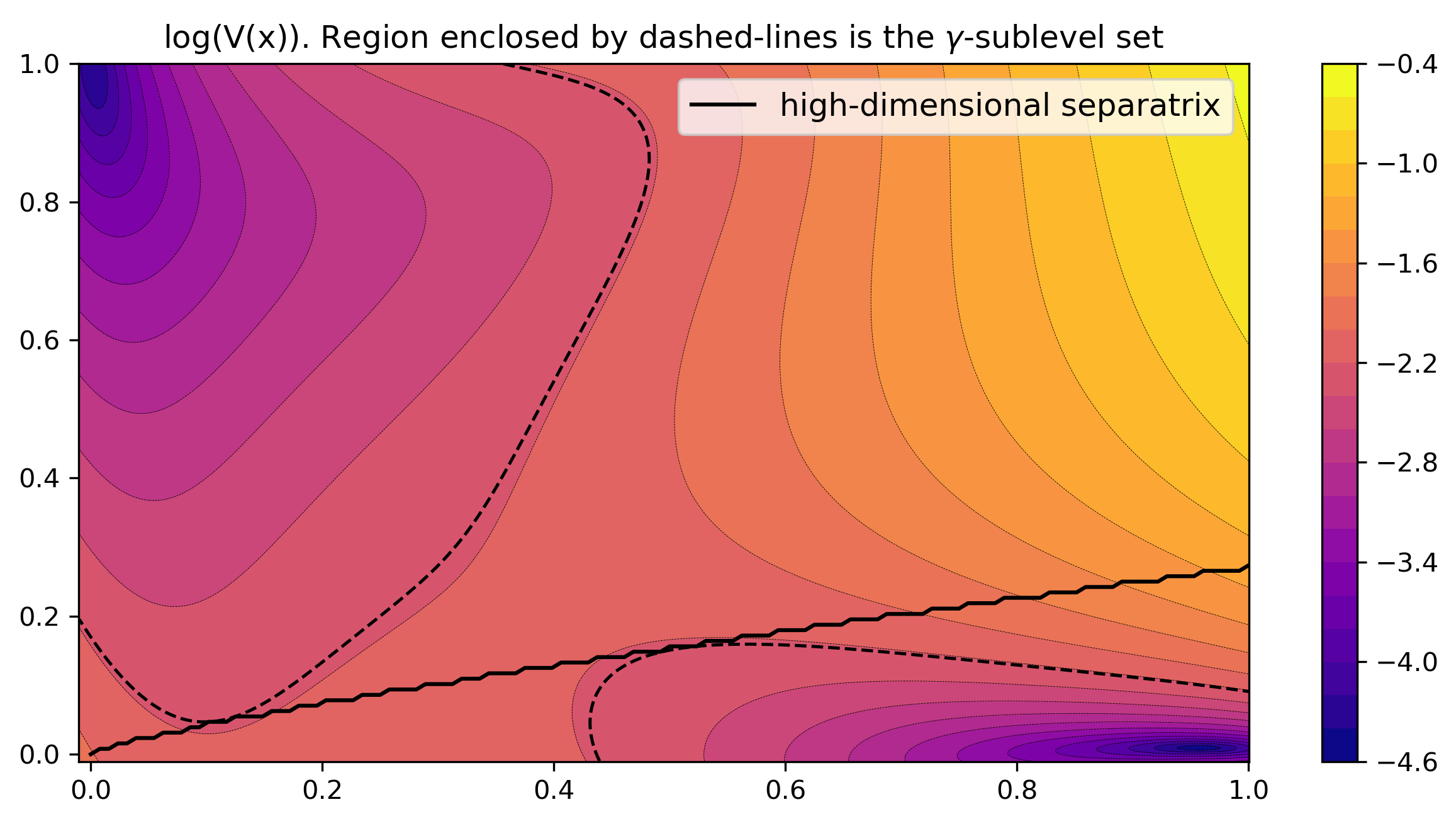}
    \caption{Two stable equilibrium points `C' and `A' (projected onto (1,0) and (0,1) respectively) and their corresponding estimated invariant sets constructed with the help of polynomial Koopman eigenfunctions of maximum degree 5. The separatrix data is incorporated into Algorithm \ref{algo:Polytope} by initializing the initial polytope with an additional constraints $\left[V_1(s_i),\cdots,V_M(s_i),-1\right]Z\ge0$ for points $s_i$ lying on the separatrix.}
    \label{fig:Invariant_set2.png}
\vspace{-0.3cm}
\end{figure}

With a more expressive Koopman basis (monomials of degree up to 5, which gives rise to a total of $4368$ terms) we apply our approach to the case of multiple stable equilibrium, wherein the invariant sets of two stable equilibrium points (labelled as `C' (i.e., ``healthy state'') and `A' (i.e., ``infected state'') in \cite{jones2019steady}) are captured by a sublevel set of the same Lyapunov function. One can leverage additional information about the system dynamics, like points lying on a manifold separating two invariant sets, as illustrated through Figure \ref{fig:Invariant_set2.png}.\\

\noindent\textit{Remarks:} SOS verification of this system was intractable, even for polynomial variable $s(x)\in\Sigma[x]$ with a small degree 4 in our SOS program. When the degree of $s(x)$ was chosen to be greater than $4$, our program terminated due to insufficient memory on a Linux machine with 32GB RAM, thus highlighting the limitations of current computational tools. We shall further explore more recent developments in SOS optimization \cite{papp2019sum,ahmadi2019dsos} in our future work on verification of Lyapunov certificates for high dimensional systems, including SMT-based verification tools that have been used extensively in recent literature, albeit for low dimensional systems \cite{formal_lyapunov,chang2019neural}. 

\section{Conclusion}\label{sec:conclusion}
Lyapunov functions are crucial for formally certifying dynamic behavior of a system and yet are not easy to construct. In many cases, it is desirable to be able to leverage trajectory data to learn these certificates. In this paper, we present a Koopman operator inspired methodology to construct Lyapunov certificates by linearly parameterizing them via basis obtained from Koopman eigenfunctions. Though this linear functional space contains Lyapunov candidates that are positive definite by construction, their time derivatives may inevitably violate the negative-definiteness condition due to limitations of data-drive learning approaches. For certifying forward-Invariance, this negative definiteness condition can be relaxed, and we present an algorithm to efficiently obtain Lyapunov functions that certify this property. We demonstrate the use of neural networks to construct our Koopman functions, but other function approximators may be used in our approach as well. In certain cases, such as polynomial-based construction, we show how optimization tools can be utilized to formally verify these learned Lyapunov functions.

\bibliographystyle{ieeetr}

\appendix
\subsection{Proof of Theorem \ref{th:main}}\label{sec:AppendixA}
\begin{proof}
For each $i$, one can write
\[
\dot{V}_i(x)=\lambda_iV_i + \varepsilon_i \le \lambda_iV_i + \|\varepsilon_i\| \le V_i(\lambda_i + \kappa_iV_i) + \omega_i.
\]
If $\lambda_i^2 > 4\kappa_i\omega_i$, there is an open interval $[\underline{\gamma_i},\overline{\gamma_i}] \subset \mathbb{R}_{>0}$ where $\dot{V}_i(x) \le 0$. This implies that the $\overline{\gamma_i}$-sublevel set of $V_i(x)$ is forward invariant. Now, for a given set of $\alpha_1,\alpha_2,\cdots,\alpha_M>0$, let us define $\gamma \doteq \min_i(\alpha_i\overline{\gamma_i})$. Thus,
\begin{eqnarray*}
     \sum_{i=1}^M \alpha_iV_i(x) \le \gamma 
     \implies \alpha_i V_i(x) \le \gamma = \min_i{(\alpha_i\overline{\gamma_i})}
     \implies V_i(x) \le  \overline{\gamma_i}
\end{eqnarray*}
for every $i=1,2,\ldots,M$.
Now, let us define a positive constant $\beta \doteq \min_i (-\lambda_i)$. Then, in the set $\left\{ x \vert \; \sum_{i=1}^M\alpha_iV_i(x) \le \gamma\right\}$ we have
\begin{eqnarray*}
    \sum_{i=1}^{M} \alpha_i \dot{V}_i(x) &\le& \sum_{i=1}^{M} \alpha_i\lambda_i V_i(x) + \sum_{i=1}^{M} \alpha_i(\kappa_i V_i(x)^2 + \omega_i) \\
    &\le& \sum_{i=1}^{M} -\alpha_i\beta V_i(x) + \sum_{i=1}^{M} \alpha_i(\kappa_i \overline{\gamma_i}^2 + \omega_i)
\end{eqnarray*}
Thus, if $\gamma\beta \ge \sum_{i=1}^{M} \alpha_i(\kappa_i \overline{\gamma_i}^2 + \omega_i)$, then we have
\[
    \sum_{i=1}^{M} \alpha_i \dot{V}_i(x) \le \beta \left(\gamma - \sum_{i=1}^{M} \alpha_i V_i(x) \right),
\]
which implies the $\gamma-$sublevel set of $\sum_{i=1}^M \alpha_i V_i(x)$ is forward invariant. Note that the condition $\gamma\beta \ge \sum_{i=1}^{M} \alpha_i(\kappa_i \overline{\gamma_i}^2 + \omega_i)$ is easily satisfied if all $\lambda_i$s are sufficiently large and negative. This completes our proof.
\end{proof}

\subsection{Proof of Lemma \ref{lemma:bounded}}\label{sec:AppendixB}
\begin{proof}
By definition, we have $\varepsilon = (\nabla^\top V_i) f - \lambda_iV_i$. Also we recall that $V_i(x) = \frac{1}{2}\|\psi_i\|^2$. Thus,
\begin{gather*}
\|\varepsilon_i\| \le \|(\nabla^\top V_i)f\| + \lambda_i V_i
\le \|\psi_i\|\cdot\|(\nabla^\top\psi_i)f\| + \lambda_i\|\psi_i\|^2.
\end{gather*}
Next, for constants $\kappa_i$ and $\omega_i$, we have $\|\psi_i\|\cdot\|(\nabla^\top\psi_i)f\| + \lambda_i\|\psi_i\|^2 \le \kappa_i\|\psi_i\|^4 + \omega_i$
\begin{eqnarray*}
\iff \|\psi_i\|\cdot\|(\nabla^\top\psi_i)f\| &\le& \kappa_i\|\psi_i\|^4 - \lambda_i\|\psi_i\|^2 + \omega_i \\
&+& p^2\|\psi\|^2 - p^2\|\psi\|^2\\
\iff \|\psi_i\|\cdot\|(\nabla^\top\psi_i)f\| &\le& \left(\sqrt{\kappa_i}\|\psi_i\|^2 - \frac{p+\lambda_i}{2\sqrt{\kappa_i}}\right)^2 \\
&+& p^2\|\psi_i\|^2 + \omega_i - \frac{(p+\lambda_i)^2}{4\kappa_i}\\
\iff \|\psi_i\|\cdot\|(\nabla^\top\psi_i)f\| &\le& \left(\sqrt{\kappa_i}\|\psi_i\|^2 - \frac{p+\lambda_i}{2\sqrt{\kappa_i}}\right)^2 \\
&+& \left(p\|\psi_i\| - \frac{q}{2p}\right)^2 + q\|\psi_i\|\\
&+& \omega_i - \frac{(p+\lambda_i)^2}{4\kappa_i} - \frac{q^2}{4p^2}\\
\impliedby \|\psi_i\|\cdot\|(\nabla^\top\psi_i)f\| &\le& q\|\psi_i\| + \omega_i - \frac{(p+\lambda_i)^2}{4\kappa_i} - \frac{q^2}{4p^2}.
\end{eqnarray*}
Thus, for a constant $q>\|(\nabla^\top\psi_i)f\|$, if we pick $\omega_i$ and $\kappa_i$ satisfying $\omega_i > \frac{(p+\lambda_i)^2}{4\kappa_i} - \frac{q^2}{4p^2}$ for some $p>0$, then 
\begin{gather*}
\|\psi_i\|\cdot\|(\nabla^\top\psi_i)f\| + \lambda_i\|\psi_i\|^2 \le \kappa_i\|\psi_i\|^4 + \omega_i = \kappa_i\|V_i\|^2 + \omega_i.
\end{gather*}
This completes our proof.\end{proof}
\end{document}